%% file: integration-main.tex
\documentclass{sig-alternate}
\usepackage{verbatim}
\usepackage{subfigure}
\usepackage{epstopdf}
\usepackage{url}
\usepackage{epsfig}
\usepackage{graphics}
\usepackage{multirow}
\usepackage{amsmath,amscd,amsbsy,amssymb}
\usepackage{color}

\usepackage[ruled]{algorithm2e}
\renewcommand{\algorithmcfname}{ALGORITHM}

\newcommand{\hide}[1]{} 

\newdef{definition}{Definition}
\newdef{problem}{Problem}
\newdef{theorem}{Theorem}

\begin{document}

\title{Social Network Integration: \\Towards Constructing the Social Graph}
\numberofauthors{4}
 
\hide{
\author{
\alignauthor Yutao Zhang\\
	\affaddr{Department of Computer Science}\\
	\affaddr{Tsinghua University}\\
	\affaddr{Beijing 100084, China}\\
	\email{stack@live.cn}
\alignauthor Jie Tang\\
	\affaddr{Department of Computer Science}\\
	\affaddr{Tsinghua University}\\
	\affaddr{Beijing 100084, China}\\
	\email{jietang@tsinghua.edu.cn}
}
}
\maketitle
\input{abstract.tex}




\input{intro.tex}
\input{problem.tex}

\input{observation.tex}
\input{approach.tex}
\input{exp.tex}
\input{related.tex}
\input{conclusion.tex}

\bibliographystyle{abbrv}
\bibliography{reference}  

\end{document}

%% file: abstract.tex
\begin{abstract}
In this work, we formulate the problem of social network integration. It takes multiple observed social networks as input and returns an integrated global social graph where each node corresponds to a real person. The key challenge for social network integration is to discover the correspondences or interlinks across different social networks.

We engaged an in-depth analysis across six online social networks, twitter, livejournal, flickr, lastfm and myspace in order to address what reveals users' social identity, whether the social factors consistent across different social networks and how we can leverage these information to perform integration. 

We proposed a unified framework for the social network integration task. It crawls data from multiple social networks and further discovers accounts correspond to the same real person from the obtained networks. We use a probabilistic model to determine such correspondence, it incorporates features like the consistency of social status and social ties across different, as well as one-to-one mapping constraint and logical transitivity to jointly make the prediction. Empirical experiments verify the effectiveness of our method.
\end{abstract}

%% file: intro.tex
\section{Introduction}
\label{sec:problem}

\begin{figure}[t]
  \centering
	\includegraphics[width =6.5 cm]{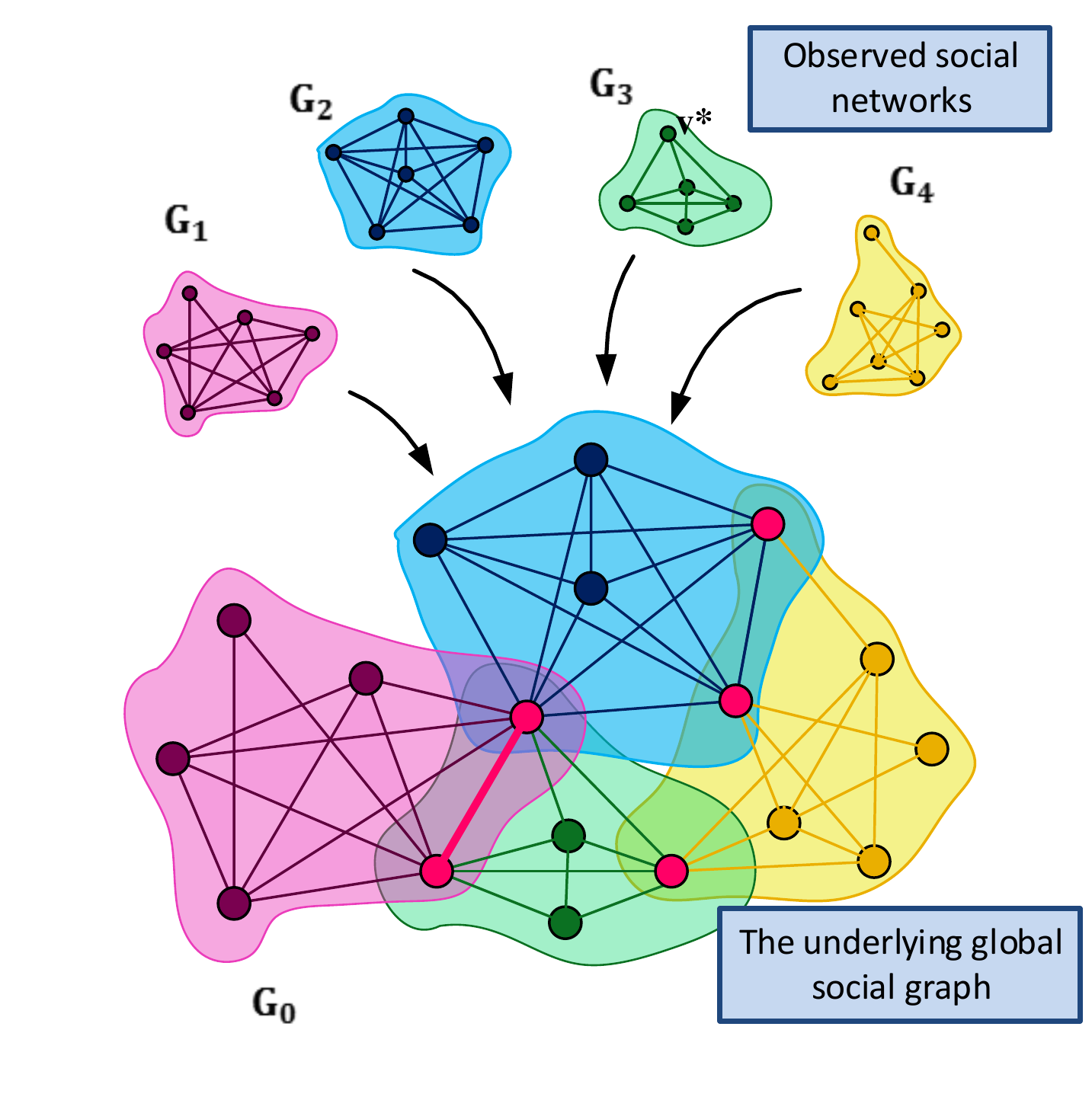}
	\caption{This figure illustrates the process of integrating four observed social networks $G_1$, $G_2$, $G_3$ and $G_4$ in order to recover the underlying social structure. Reversely, these social networks are subset of integrated global social graph. Red colored nodes are users that have accounts on multiple social networks and red edges represented co-occurrence social ties across different social networks.}
	\label{pic:intro}
\end{figure}

Online social media capture the digital trace of our social life. The diverse functionality of different on-line social media services make us join multiple social networks. We stay in touch with our friends on facebook, follow celebrities on microbloging sites and manage business contacts through linkedin. As a result, our social life has been broken into pieces due to the absent of the interlink between each social network. Is it possible to connect these pieces together, in order to recover the full image of our social life?

The advantage of such an process is multi-folded. For an individual user, managing multiple social network accounts takes a lot of effort. Bridging each isolated social circles of a user could lighten the communication cost and reduce information redundancy. From a service provider's perspective, aggregating information from different social networks gives a more comprehensive view of each individual user and filters out the potential bias, hence it would benefit many applications. Taking recommendation system as an example, it would be a straightforward solution to cold start problem by leveraging information from the accounts on other sites of a new user. Moreover the inter-connections between different social networks might reveal interesting yet unobservable phenomenons when we look at an individual social network independently, such as the information diffusion pattern across different social media. However, this could also lead to potential leakage of privacy as it exposes users' identity without permission. 

In this work, we are trying to tackle the problem of social networks integration, where integration means combining the accounts correspond to the same real person across different social networks, in order to get a more complete view of that person's social life. We want to investigate whether, how and to which extent we can fulfill this task. 

Figure \ref{pic:intro} gives a toy example illustrating the process of integrating four observed social networks $G_1$, $G_2$, $G_3$ and $G_4$ in order to recover the structure of the underlying global social graph $G_0$. The upper side of Figure \ref{pic:intro} is the input of our problem: observed social networks. Nodes represent accounts and edges represent social ties within each individual social network. At the bottom shows the output of the problem: the global social graph integrates the four observed social networks. Users who have accounts on multiple social networks are colored with red. Such an intuitive process is non-trivial due to the following reasons. Firstly, lacking of unique identifier. There is no such an attribute that can uniquely identifies a person across different social media, even name won't satisfy the condition since a single name can be shared by different person and also can be represented in different ways. Secondly, the heterogeneity of different social networks, where the heterogeneity reflected in both topology and semantics on different social networks. As a result, the features of an individual user are not necessarily be consistent across different social networks. The third challenge comes from data availableness, because of the privacy policy and the large and the on growing data scale, we can hardly observe the entire network. 

Considering the above intuitions, in the paper, we propose a social network integration framework which effectively integrates arbitrary number of social networks. In the framework, the data is obtained a set of ad-hoc crawlers each corresponds to an individual social networks. A probabilistic model is used to discover the accounts on each network that corresponds to the same real person, hence integrates the obtained social network. The result of integration will be further feedback to the crawlers to adjust crawling priority.

In summary, our contributions in this work include:
 \begin{itemize}
 \item We formulate the problem of social networks integration, and discuss associated challenges and potential;
 \item We propose a unified social network integration framework that takes multiple social networks as input, and returns a integrated social graph.
 \item Empirical experiments on integrating three large scale social networks, Linkedin, Videolectures and ArnetMiner demonstrate the effectiveness of our framework.
 \end{itemize}

The rest of the paper is organized as follows. Section 2 formally defines the problem of social network integration; Section 3 gives an in-depth analysis on the three social networks; Section 4 is devoted to our social network integration framework; Section 6 analyzes experimental results; Section 7 presents related works and Section 8 concludes this work and discuss future directions.

%% file: problem.tex
\section{Problem Definition}
\label{sec:problem}

In this section, we give several necessary definitions and present the formulation of social network integration problem.

Given $k$ observed social network $G = \{G_1, G_2, ..., G_k\}$. Each social network is an unweighted undirected graph $G_i = (V_i, E_i)$, where $V_i = \{v_i^1, ..., v_i^{|V_i|}\}$ is a set of social network accounts and each edge $(v_i^{k_1}, v_i^{k_2}) \in E_i$ represents a social tie or social interaction between two accounts $v_i^{k_1}$ and $v_i^{k_2}$ within social network $G_i$. Social network integration problem can be divided into the following two sub-tasks:\\

{\bf Probabilistic Alignment.} An \emph{alignment} $O$ is a set of triplets  $o^k_{i,j} =(v_i,v_j,p_k)$ each denotes an assertion that $v_i$ from network $G_i$ and $v_j$ from network $G_j$ potentially belongs to the same natural person, where $p_k$ is a real valued probability indicates the confidence. The aim of probabilistic alignment task is to discover such inter-network correspondence within the observed social networks $G$.\\

{\bf Social Graph Construction.} With the alignment generated at the last step, we further constructing the social graph, where \emph{global social graph} $G_0(V_0, E_0)$ is the underlying global network structure given a set of observed social network $G$. It is represented as a multi-graph, where each node $v_0^k \in V_0$ is a set of node in different networks that corresponds to an individual natural person. An edge $e_0^{k_1,k_2}$ between $v_0^{k_1}$ and $v_0^{k_2}$ will be created if there is an edge between two nodes $v_i^{k_1} \in v_0^k$ and $v_i^{k_2} \in v_0^{k'}$ in any social network $G_i$. Given social networks set $G = \{G_i\}_i^k$, our goal is to reconstruct the underlying social graph $G_0$ by integrating all the social networks in $G$. For any node $v_0^k$ in $G_0$, we know exactly the set of $\{v_i^{k'} \in v_0^k \}$.

\hide{

As mentioned in Section 2, the key to network integration lies in the finding of alignment $O_{i,j}$ between any two given social network $G_i \in G$ and $G_j \in G $, which determines the structure of the global social graph. Directly searching for the optimal $O_{i,j}$ would be difficult since scale of the search space is $O(|V_i| \cdot |V_j|)$ for any two social network $G_i$ and $G_j$. Thus we first generate candidate pairs $C = \{c| c \in C(i,j),  i \neq j \wedge i, j = 1,2,...K\}$, where $O \subseteqq C$. Candidates of node $v_i$ in $G_j$ is denoted as $C_j(v_i)$. Section 5 illustrates how to generate candidate pairs effectively, and Section 6 explains how to find $O$ accurately. Symbols and notations are summarized in Table \ref{table:notation}.

is a set of node in $\bigcup_0^k{V_i}$ that are corresponding pairs with each other.. An edge $e_0^{k_1,k_2}$ between $v_0^{k_1}$ and $v_0^{k_2}$ will be created if there exists at least one edge between $v_i^{k_1} \in v_0^k$ and $v_i^{k_2} \in v_0^{k'}$ in $G_i$.}

\begin{table}[t]\centering
\begin{tabular}{|c|c|}
\hline
$G$ & A set of  social networks $G_1$, $G_2$, ... , $G_K$\\
\hline
$G_0$ & The global social graph given $G$\\
\hline
$v^k_i$ & A given node from $G_i$\\
\hline
$C_{i,j}$ & The candidate set between $G_i$ and $G_j$\\
\hline
$c_{i,j}^k$ & A given candidate pair in $C_{i,j}$\\
\hline
$O_{i,j}$ & The alignment of $G_i$ and $G_j$\\
\hline
$o_{i,j}^k$ & A given corresponding pair between $G_i$ and $G_j$\\
\hline
\end{tabular}\caption{Symbols and Notations}\label{table:notation}
\end{table}

%% file: observation.tex
\section{Data and Observations}
\label{sec:observe}
\subsection{Data Collections}
As we addressed in Section 1, to perform network integration, one of the challenges comes from data availableness. Obtaining data for social network integration is not an easy task as in most cases, we won't be able to observe the complete networks due to the size and privacy policy. In this work, we collected data from three social networks: Linkedin\footnote{http://www.linkedin.com/}, VideoLectures\footnote{http://www.videolectures.org/} and ArnetMiner\footnote{http://www.arnetminer.org/}. \\

{\bf Linkedin's ``Also-View'' network:} 
Linkedin operates the world's largest professional network on the Internet. It has more then 238 million members in over 200 countries and territories\footnote[4]{http://press.linkedin.om/about}. On Linkedin network, part of user information is available for public profiles, such as education, affiliation and industry(refer to Figure 5). In addition, users could maintain a certain amount of connections, which could be friends, co-workers, etc. However, personal connections are viewed as private information and unavailable to public access. But we discovered through the ``Viewers of this profile also viewed... '' table at the right side of public profiles. Hence, we construct an undirected network based on the "also-view" links, which can be considered as a k-nearest neighbor network. We crawled approximately 3 million public profiles and more than 25 million ``also-view'' relationships.\\

{\bf Videolectures' ``Together with'' network:}
Videolecture is an open access educational video lectures repository. The lectures are given by distinguished scholars and scientists at most important events like conference, summer schools and workshops\footnote[6]{http://videolectures.net/site/about/}. Connections between 11777 profiles we crawled are not explicitly provided by Videolecture. However, researcher attending the same venue can view as a kind of social relationship or interaction. Hence we extracted 786353 ``Together with'' relationships out of all profiles.
In Videolecture's ``together with'' network, an undirected edge $e = (v_i, v_j)$ would be created if $v_i$ and $v_j$ have at least co-attended a venue.\\

{\bf ArnetMiner's co-author network:} 
Arnetminer aims to provide comprehensive search and mining services for researcher social networks. We obtained the entire ArnetMiner data up to year 2013, including user profiles and co-author relationships. The details of user profiles ontology are shown in Figure 5. To construct co-author network, we create a undirected edge $e = (v_i, v_j)$ if author $v_i$ and $v_j$ have coauthored a publication. Finally, we obtain a co-author network with more than 1 million nodes and nearly 4 million edges.\\

 The statistics of the three networks are illustrated in Table \ref{table:data}.\\
\begin{table}\centering
\begin{tabular}{|c|c|c|}
\hline
Observed social network & \# node & \# edge \\
\hline\hline
Linkedin & 2985414 & 25965384\\
ArnetMiner &1053188 & 3916907\\
Videolecture&11178 & 786353\\
\hline
\end{tabular}
\caption{Data statistics}
\label{table:data}
\end{table}

{\bf Crowdsourcing data labeling:}
To construct training set and evaluate our method, we need ground truth of alignments between different social networks. Labeling such data is a time-consuming and labor intensive work. It requires human annotator to manually searching for the accounts belong to a certain person on different social networking service. 

\begin{table}\centering
	\begin{tabular}{|c|ccc|}
	\hline
	Labeled Pairs & $|L(A,*)|$ & $|L(L, *)|$ & $|L(V,*)|$\\
	\hline\hline
	Aminer &-&7532&1470\\
	\hline
	Linkedin &7532&- &410\\
	\hline
    Videolecture &1470&410&-\\
    \hline
	\end{tabular}
 	\caption{Statistics of labeled pairs between different social networks, where $|L(*,*)|$ denotes the number of labeled corresponding pairs between the given two networks.}\label{table:ground-truth}
\end{table}

We leveraged crowdsourcing technique to ease the labeling work. An editable column has been provided on academic search service we developed for user to fill in the urls link to the external social network accounts of the corresponding author. This service has been running on-line for more than one year and more than 10,000 interlinks record has been collected from the users. We manually checked all the records and filtered out those are invalid or redundant. Finally, about 7500 interlink between ArnetMiner network and Linkedin network has been obtained. The statistic of labeled ground truth are provided in Table \ref{table:ground-truth}.

\subsection{Analysis}
As the first step, we engage in some high-level investigation of the generality and particularity of different social networks. Hence, we want to answer two questions:
 \begin{itemize}
 \item What reveals our social identity?
 \item What is consistent across different social networks?
 \end{itemize}

\subsubsection{What reveals our social identity?}
\begin{figure}\centering
\subfigure[ArnetMiner]{
\includegraphics[width = 3.5 cm]{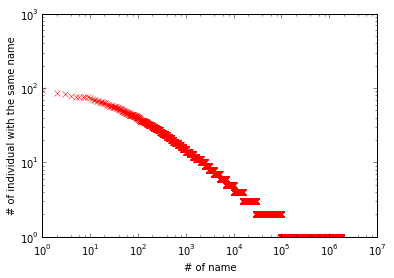}
}
\subfigure[Linkedin]{
\includegraphics[width = 3.5 cm]{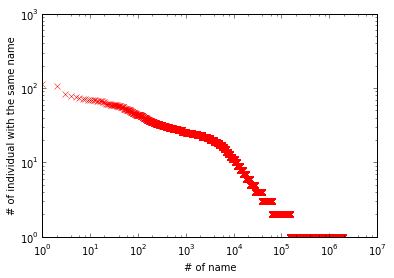}
}
\subfigure[Videolectures]{
\includegraphics[width = 3.5 cm]{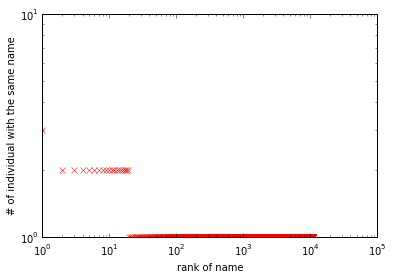}
}
\subfigure[All together]{
\includegraphics[width = 3.5 cm]{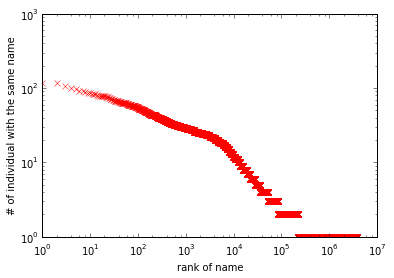}
}
\caption{Number of person sharing an individual username on each social network}
\label{duplicate}
\end{figure}

{\bf Name:}
Username is the identifier of a user on social media sites. We make the assumption that users tend to use similar user name in different social networks. Such an intuition becomes more obvious since we are dealing with real name social networks in this work. However, even if in real name social networks, name wouldn't necessarily gives away user's identity since: 1) A name can be shared by different person. Figure \ref{duplicate} shows the statistics of account sharing an individual name in each social network, it is not surprising to see a few names are shared by more than 100 accounts. While for Videolectures network, there is not as much name duplication since the data scale is smaller. 2) A name can be represented in different ways, as in ArnetMiner, many authors' first name only contains the initial letter . One intuition is that rare used names are likely owned by a single real person. For example, given two accounts both named "Wei Wang", we wouldn't have much confidence to claim they belong to the same user since there are so many people using this name. On the contrary, if two accounts share the name "Francis Scott Key Fitzgerald", it is safe for us to make that assumption. So in a word, we can't simply integrate accounts with the same name, but at least it gives us some clue.

\begin{figure}
  \centering
  \includegraphics[width=7.5 cm]{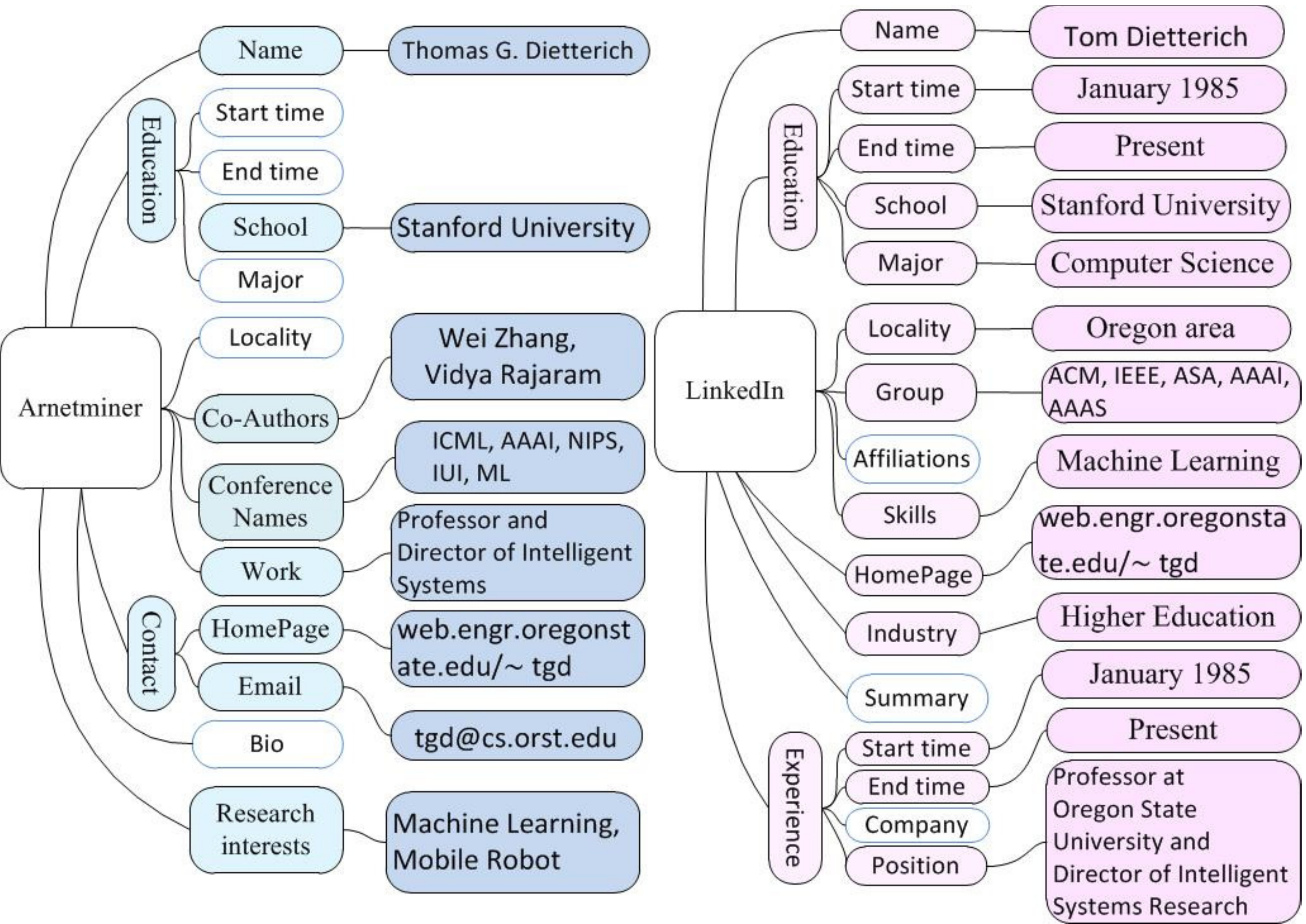}\\
  \caption{Heterogeneous profile ontology of ArnetMiner (left) and Linkedin (right).}\label{pic:profile}
\end{figure}

{\bf Profile:}
Profile information is the description of a person given by itself, which is very informative. Unfortunately, it's not easy to directly use due to the following two reasons. Firstly, profile ontology on different social networks are heterogeneous. Figure \ref{pic:profile} illustrates the profile structure of ArnetMiner and Linkedin. Both ArnetMiner and Linkedin have name, education and locality, but the rest are different. However, in most of the cases, group and skills in Linkedin profile could be mapped to research interests in ArnetMiner; and summary has similar contents in Bio. In addition, homepage in Linkedin is sometimes contained in ArnetMiner's contact, and the same applies to industry (in Linkedin) and work (in ArnetMiner). Hence, information is unstructured and distributed. As for Videolectures, we only have name, affiliation and homepage. As a result, the heterogeneous nature of different social networks makes no guarantee of similar profile structure or structured information, which poses a big challenge to our algorithm design. Secondly, user provided profiles are noisy and incomplete. According to our statistics,  more than half of the Linkedin users do not have summary, interests, specialties, etc, such a problem is even more serious on ArnetMiner network. 
In user profiles, some fields can uniquely identify a person, such as Email, homepage url. These information are not commonly available but once we discover two accounts share such kind of attribute, we can safely conclude that they correspond to the same person.

{\bf Friends:}
"Friends make us who we are" \cite{flora2013friendfluence}. It is possible that the social contacts we created within the social networks  expose our identity. We have already mentioned the existence of consistent social ties that co-occurs on different social network. What we want to address here is that different friend reveals our identity to varying degree. Which is quite straight forward since when someone introduced himself as a fan of lady gaga, it almost gives away nothing about his social identity while if he mentioned he is a friend of someone not so famous, at least his identification has been narrowed down to a finite set of person. Further, given two accounts $v_i^{k_1}$ and $v_j^{k_2}$, if we already known they share a set of friends, can we claim they belongs to the same person? An obvious intuition is that the more shared friends they have, the more likely the accounts belongs to one person, but it's not the only factor. Imagine if a person is connecting with a strong connected component, according to social balance theory \cite{}, the probability of existing another person connecting with that strong connected component is clearly larger than connecting a bunch of person scattered in different communities. Such a intuition can be further confirm by sub-graph frequency of networks\cite{ugander2013subgraph}. 

\subsubsection{The consistency across different social networks}

\begin{figure}[t]\centering
\includegraphics[width = 7 cm]{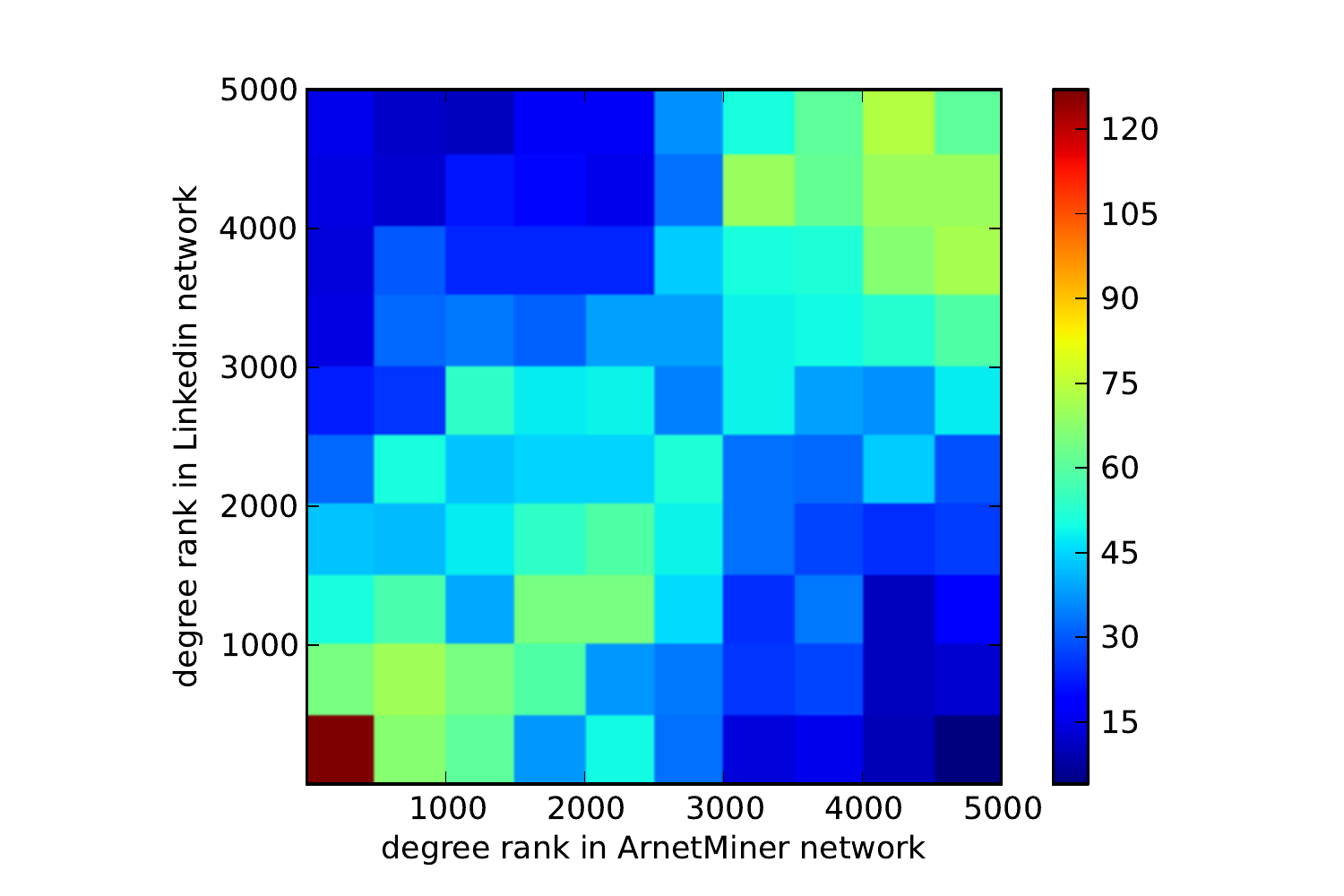}
\caption{Users' social status in Linkedin network and ArnetMiner network}
\label{pic:social-status}
\end{figure}

{\bf Social Status:}
In sociology, social status is the position or rank of a person or group within a social system. Social status indicates the influential ability and the role a user plays in the social network. As an important social factor on every social network, we want to see whether the social status of an individual person is consistent on different social systems. For simplicity, we use degree as the measurement of social status of a user, which suggests that the person with more friends has the higher social status. We ranked the labeled pairs between ArnetMiner network and Linkdin network by it's degree on the two network respectively and plotted on a two dimensional space. In Figure \ref{pic:social-status}, X-axis indicates the rank on ArnetMiner and Y-axis for Linkedin, the color indicates the density of each region. 

From Figure \ref{pic:social-status} we can see that data points tend to appear near the diagonal which suggests to some degree, the social status can be considered as consistent. In addition, we can clearly see the left bottom corner is the densest region which shows the social status of opinion leaders are more consistent than others. This observation is very intuitive since building reputation and becoming an opinion leader on a social network is an expansive process while being an opinion leader on a social network shows the users have enough ability and social capital, therefore he/she is more likely to also be a opinion leader on a different social network.

{\bf Social Ties:}
The semantic of relationships or social ties within different social networks are interpreted differently. As we have addressed in 3.1, the edge in ArnetMiner network represents co-authorship while it is interpreted as business contact or coworker relationship in Linkedin, and in Videolectures network, it represents two person co-attending a venue. Although the meaning of them are different, intuitively we believe that there will be co-occurrence of social ties across different social networks. The explanation is straight forward, the social tie within a network implies two persons already know each other, so they are more likely to create a relationship in another social network than two random persons. 
\begin{figure}[t]\centering
\includegraphics[width = 7 cm]{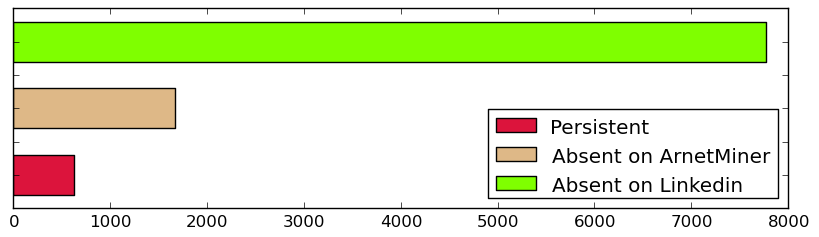}
\caption{Persistence of social tie between Linkedin network and ArnetMiner network}
\label{pic:social-tie}
\end{figure}
We investigate the consistency of social tie on the labeled corresponding pairs between ArnetMiner network and Linkedin network. Figure \ref{pic:social-tie} shows the number of co-occurred social tie between these users. Here, we say a social tie is co-occurred if it exists on both social networks. There are 627 co-occurred social ties, while 1670 appear on Linkedin but absent on ArnetMiner and 7773 conversely. Note that we only partially observed the Linkedin network due to the privacy protection, hence, we can safely assume that there will be much more co-occurred social tie if the complete network is available.

%% file: approach.tex
\section{network integration pipeline}
\label{sec:framework}
Base on the observations above, we proposed a unified framework to integrate multiple observed social networks in order to recover the underlying structure of the global social graph.
Figure \ref{pic:PwCG} shows overview workflow of our social network integration framework. The profile accounts crawled from different social networks will be first converted to candidate pairs. Then, the candidate pairs will be feed into a probabilistic model in order to determine the correctness of each candidate pair.

\subsection{Candidate Pairs Generation}
Finding pairs of accounts on two social networks $G_i(V_i, E_i)$ and $G_j(V_j, E_j)$ that belong to the same natural person, the scale of search space is $O(|V_i|\cdot |V_j|)$, and even larger for integrating more than two social networks. Such a task is intractable for real world networks with millions of nodes. It is obvious that only very few pairs of accounts are actually belong to the same person, hence we need to quickly and roughly filter out those pairs that are unlikely to be true. Therefore, we use a candidate pairs generation step to reduce the search space. A desired candidate generation method should have following characteristics: 
 \begin{itemize}
 \item Easy to implement
 \item Scalable to very large networks.
 \item Generate small candidate sets meanwhile covers most of the corresponding pairs.
 \end{itemize}
As in , we make the assumption that user tend to use similar user name in different social networks. Such an intuition becomes more obvious since we are dealing with real name social networks in this work. Here we take this heuristic to generate candidate pairs that are potentially belong to an individual natural person.
 
To tolerate miss spell and different representation of names, we decompose each name into segments and group the accounts sharing at least one segment in the name. To further shrink the size of candidate sets, we compute jaro-winkler distance between the name of accounts within a group. Finally, an appropriate threshold is chosen to trade off between candidate size and coverage rate. Section 6.1 presents empirical analysis on different thresholds.
\\
As we can see, our method is easy to implement; scalable to large social networks; and as showing in Section 6.1, it yields high coverage rate of corresponding pairs. In addition, this method fits online streaming data well. Candidate sets could be effectively updated incrementally with the incoming data.

\subsection{Learning To Align Corresponding Pairs}
With the candidate pairs generated at last step, we need to further determine which of them are truly corresponding pairs, which is a binary classification task to assign the true or false label for each candidate pair. 

\begin{figure}
  \centering
  \includegraphics[width=7 cm]{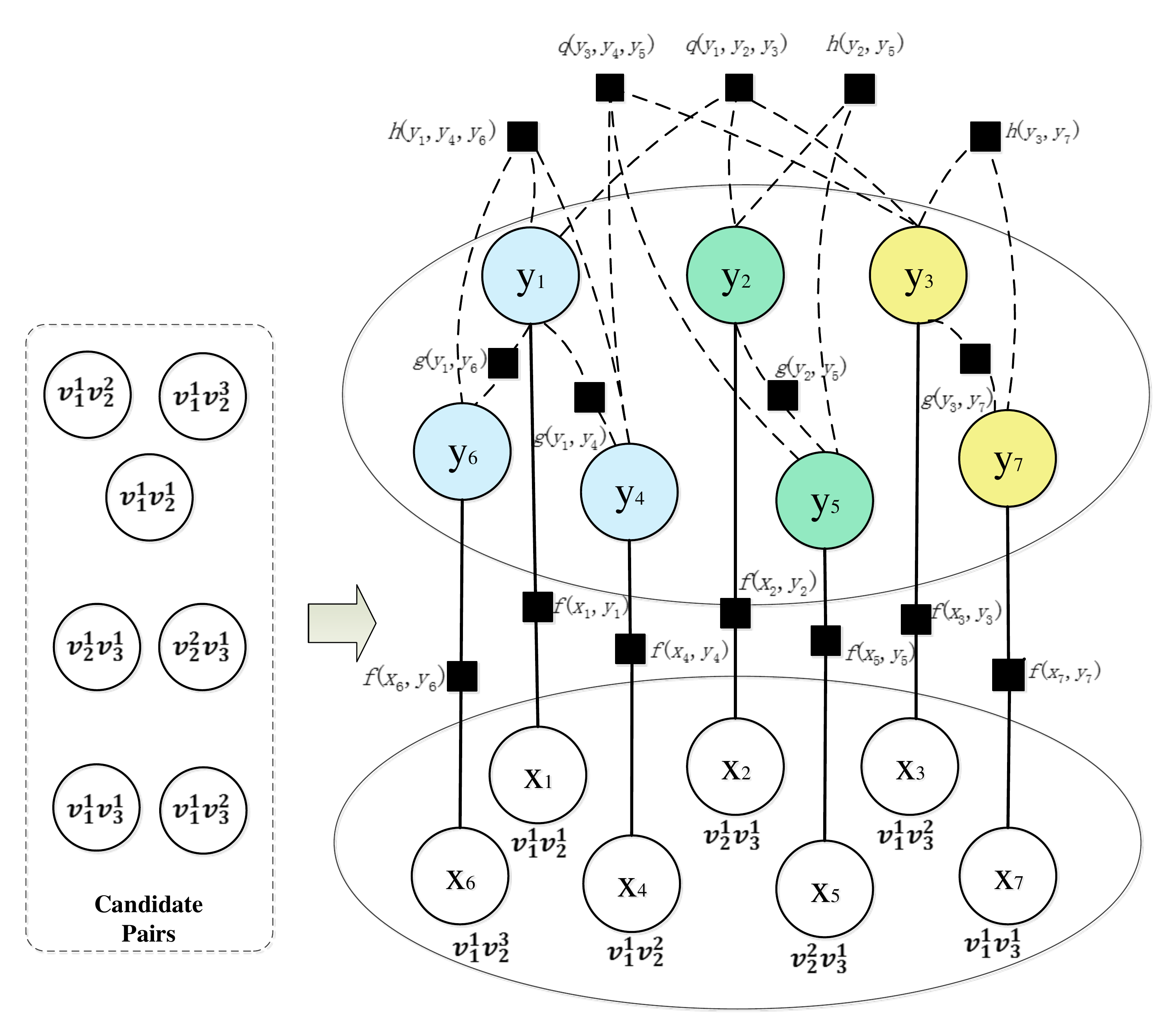}\\
  \caption{Graphical representation of the corresponding pairs alignment model}\label{pic:PwFG}
\end{figure}

We formulate the classification problem into a probabilistic graphical model in order to capture the structural correlations and the uncertainty in predictions. Each candidate pair $c^k \in \cup C_{i,j}$ is mapped to an observed variable $x_k$. $Y = \{y_k\}_{k=1}^{|\cup C_{i,j}|}$ are hidden variables that represents the true or false labels of $X = \{x_k\}_{i=k}^{|\cup C_{i,j}|}$. To better incorporate the observations discussed in Section 3.3, we use a factor graph to represent the probabilistic model. We defined two types of feature functions:

\begin{itemize}

\item Local feature function: $f(y_k, x_k)$ captures the characteristics between the two accounts $v_i$ and $v_j$ associated with the latent variable $y_k$. 

\item Correlation feature function: $g(y_k, y_{k'}, \mathcal{G}(y_k,y_{k'}))$ denotes the correlation between two latent variables $y_k$ and $y_{k'}$.

\item Group constraint function: $h(y_k, \mathcal{H}(y_k, \mathcal{C}_k))$ represents the correlation between a latent variable $y_k$ and a constrained group $\mathcal{C}_k$. 

\item Triangle feature function: $q(y_k, y_{k'}, y_{k''}, \mathcal{Q}(y_k, y_{k'}, y_{k''}))$ denotes the correlation within the three latent variables $y_k$, $y_{k'}$ and $y_{k''}))$ form a triangle closure.

\end{itemize}

Figure \ref{pic:PwFG} gives a graphical representation of the proposed model. Given the observed social networks $G$, candidate pair set $C=\cup{C_{i,j}}$, we can define the joint distribution over $Y$ as:
\begin{equation}
\begin{aligned}
p(Y|G, C) = & \prod_k f(x_k,y_k)g(y_k, \mathcal{G}(y_k, y_{k'})) \\
            & h(y_k, \mathcal{H}(y_k, \mathcal{C}_k))q(y_k, \mathcal{Q}(y_k, y_{k'}, y_{k''}))
\end{aligned}
\end{equation}

\subsection{Feature Design}

{\bf Modeling the Uniqueness of Name:}
As we mentioned in 3.3.1, an unique name are likely to be own by a single natural user. To measure the uniqueness of a name, we use a language model based approach. Given a name $\mathcal{N} = (s_1,s_2,...,s_n)$, where $s_k$ is the segment obtained at candidate generation step. We define uniqueness of a name as $\mathcal{U}(\mathcal{N}) = -\sum_{k=1}^n {log {P(s_k)}}$. For each candidate pair $c_{i,j}^k = (v_i^{k_1},v_j^{k_2})$, a feature $\mathcal{U}(c_{i,j}^k) =(\mathcal{U}(\mathcal{N}(v_i^{k_1}))+\mathcal{U}( \mathcal{N}(v_j^{k_2}))) / 2$ is assigned to capture whether the two accounts are sharing a unique name.

{\bf Modeling Profile Similarity:}
 Hence, we categorized user profile fields into \emph{Strong Identifier} and \emph{Weak Identifier}. We calculate the edit distance between each shared strong identifier for a given candidate pairs as a set of features. For weak identifiers, we concatenate them into a single document and convert each document into a bag-of-words vector. For each vector, the words are weighted by TF-IDF. The similarity between the profiles of each candidate pairs are measured by inner product and cosine distance.

{\bf Modeling Social Status Shift:}
Based on the observation in Section 3.3.1, the pattern of social status shifting across different social networks can give us discriminative power to determine true or false label of candidate pairs. To incorporate this observation into our predictive model, we first determine the social status of each account within it's social network, again, we use degree to measure social status. We sorted accounts by degree in each social networks and we call the top $1\%$ accounts "opinion leaders", the following $10\%$ "middle class", and for the rests, we call them "the messes". Given a candidate pair $c_{i,j}^k = (v_i^{k_1}, v_j^{k_2})$, where $v_i^{k_1}$ is an opinion leader in $G_i$ and $v_j^{k_2}$ is a middle class in $G_j$, a feature $from O in G_i to M in G_j$ will be assigned to it.\\

\begin{figure}[t]
  \centering
  \includegraphics[width=6.5 cm]{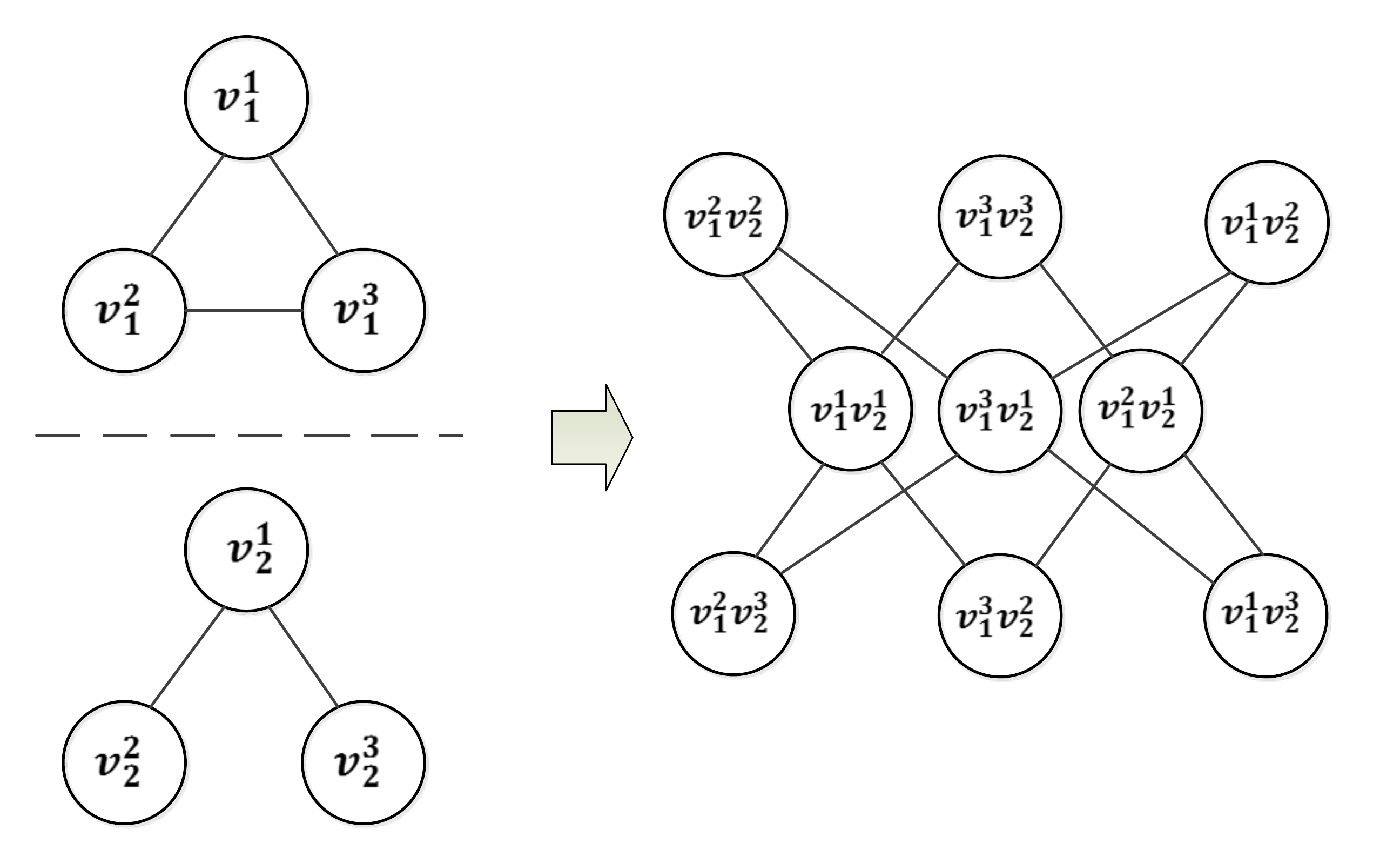}\\
  \caption{Modeling co-occurred social tie across social networks $G_1$ and $G_2$}\label{pic:PwCG}
\end{figure}
\begin{figure}[t]
  \centering
  \includegraphics[width=6.5 cm]{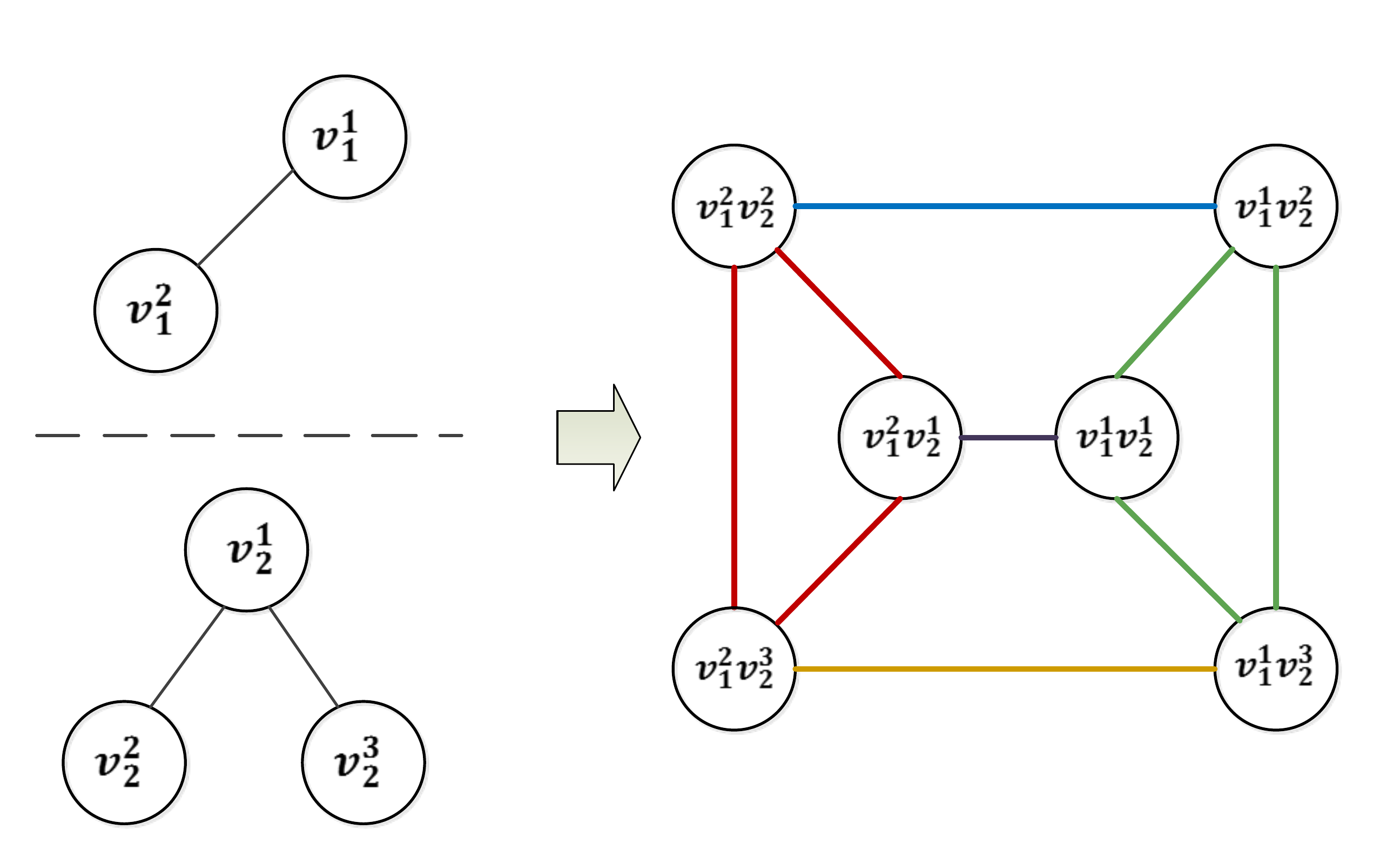}\\
  \caption{Modeling one-to-one mapping constrain between social networks $G_1$ and $G_2$}\label{pic:constrain}
\end{figure}
\begin{figure}[t]
  \centering
  \includegraphics[width=6.5 cm]{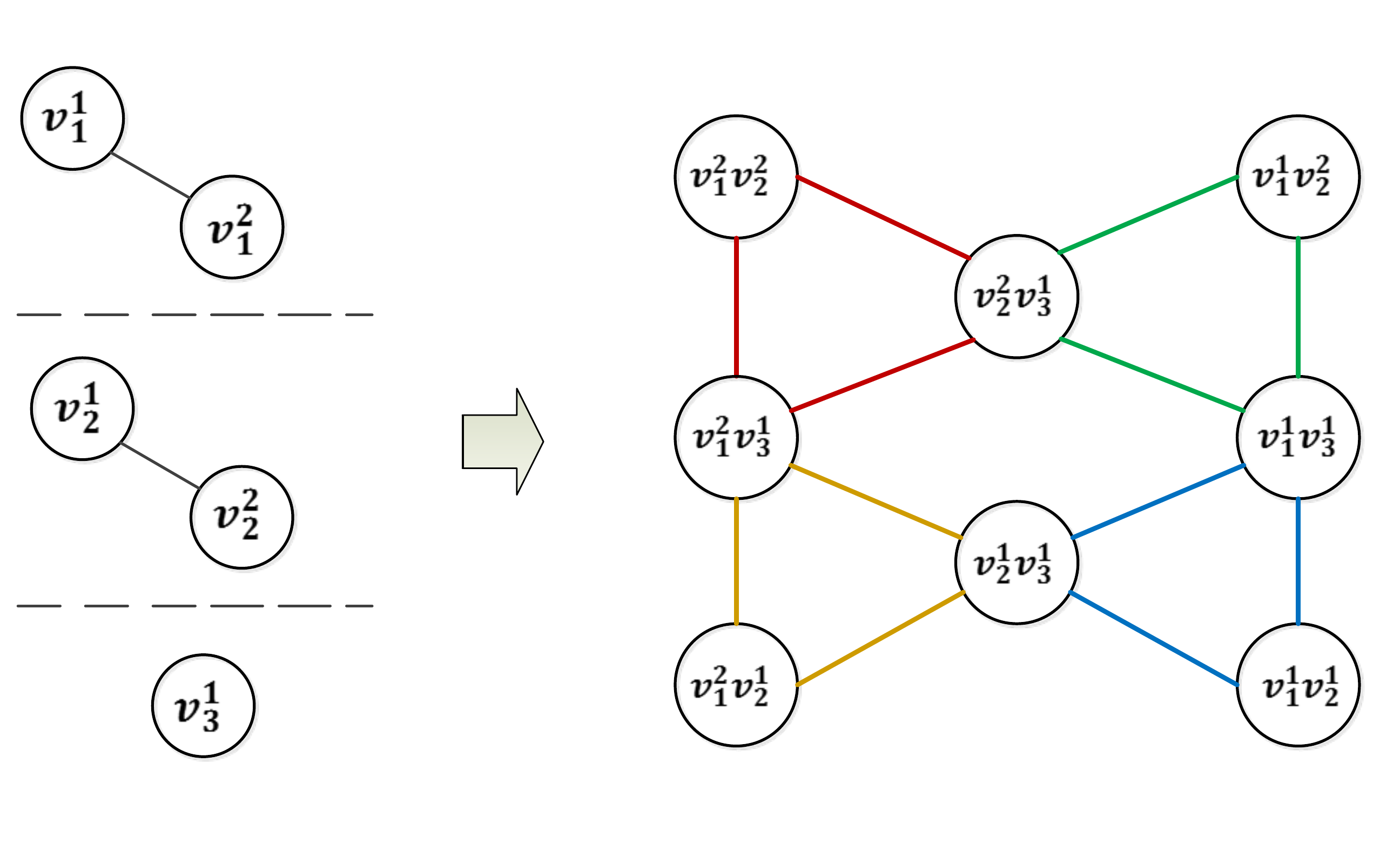}\\
  \caption{Modeling logical transitivity on integrating three social networks $G_1$, $G_2$ and $G_3$}\label{pic:transitivity}
\end{figure}

The features defined above capture the characteristics between within a candidate pairs, hence, we incorporate them as local features, which is defined as an exponential function:
\begin{equation}
f(y_k, x_k) = \frac{1}{Z_{\alpha}} e^{\sum_{d}{\alpha_d\Phi_d(y_k, x_k)}}
\end{equation}
Where $\Phi_d(.)$ is the $d^{th}$ local feature defined on $x_k$ with respect to the value of $y_k$; $\alpha_d$ is the weight of the feature; $Z_{\alpha}$ is a normalization factor.\\

{\bf Modeling Co-occurred Social Ties:}
According to Section 3.3.1, a social tie may co-occur on different social networks if both of the two associated users have accounts on those networks. However such a strong signal is not straight forward to capture since we do not know the correspondence between social ties before identifying the users associated with them. Here, we model the co-occurred social ties as the structural correlation between candidate pairs. To make it more intuitive, Figure \ref{pic:PwCG} gives a toy example illustrating this process. On the left side are two networks $G_1$ and $G_2$. On the right side, each node represents a candidate pair, and each edge represents a potential co-occurred social tie. In the example, we assume any two node from the two network is a candidate pair, hence there are $3 \times 3 = 9$ nodes on the right. We construct the potential co-occurred social ties set $\mathcal{CST}$ with the following process. For each two candidate pairs $c^{k} = (v_i^{k_1}, v_j^{k_2})$ and $c^{k'} = (v_i^{{k_1}'}, v_j^{{k_2}'})$, we draw an edge between them if there is an edge $e_i^{k_1,{{k_1}'}} = (v_i^{k_1}, v_i^{{k_1}'})$ in $G_i$ and an edge $e_j^{k_2,{{k_2}'}} = (v_j^{k_2}, v_j^{{k_2}'})$ in $G_j$ respectively, we say $e_i^{k_1,{{k_1}'}}$ and $e_j^{k_2,{{k_2}'}}$ are potential co-occurred social ties between $G_i$ and $G_j$ and there is correlation between $c^{k}$ and $c^{k'}$.

Co-occurred social tie is formulated as a correlation feature which is defined as:
\begin{equation}
  g(y_k, y_{k'}, \mathcal{G}(y_k,y_{k'}))  = \frac{1}{Z_{\beta}} e^{{\beta\textbf{g}(y_k, y_{k'})}}
\end{equation}
$\mathcal(y_k, y_{k'})$ is an indicator to specify whether there is a potential co-occurred social tie between two candidate pairs, thus, $\textbf{g}(y_k, y_k) = 1$ if there is, and 0 otherwise.

{\bf Modeling One-to-one Mapping Constraint:}
Users may create alias accounts on social networks, while in most cases users will stick to a single account since managing one account is hard enough. Hence we set the constraint that one-to-one mapping is encouraged between two networks. Formally, when integrating network $G_i$ and $G_j$, we say a set of candidate pairs are constrained by $v_i^k$ if all of these candidate pairs contains node $v_i^k$ since only one of them are likely to be true, and we call it a constrained group $\mathcal{C}_k$. Figure \label{pic:constrain} shows the one-to-one mapping constraint when integrating networks $G_1$ and $G_2$. As an example, the three candidate pairs $(v_1^2,v_2^2)$, $(v_1^2,v_2^1)$ and $(v_1^2,v_2^3)$ connecting by red edges are constrained by $v_1^2$ since all of them are associated with $v_1^2$. We use $\mathcal{MC}=\{\mathcal{C}_k\}$ to denote all the one-to-one mapping constrained group.

One-to-one mapping constraint feature is defined as
\begin{equation}
h(y_k, \mathcal{H}(y_k, \mathcal{C}_l)) =  \frac{1}{Z_{\gamma}} e^{{\gamma\textbf{h}(y_k, \mathcal{C}_l)}}
\end{equation}
where ${C}_l \in \mathcal{MC}$ is a constrained group that $y_k$ in. $\textbf{h}(y_k, \mathcal{C}_l)$ is set to 0 if more than two of variables in $\mathcal{C}_l$ labeled true, since it violates the one-to-one mapping constraint, and 1 otherwise.

{\bf Modeling Logical Transitivity:}
Since the assertion of account A and account B belong to the same user is symmetrical, there is logical transitivity between related mapping. Figure \ref{pic:transitivity} shows the logical transitivity when integrating three social networks $G_1$, $G_2$ and $G_3$, where each transitivity closure is indicated by an individual color. For instance, the three candidate pairs $(v_1^2,v_2^2)$, $(v_1^2,v_3^1)$ and $(v_2^2,v_3^1)$ forms a transitivity closure, and constrained by the fact that if two of them are labeled true, the other one must be true. Hence when we dealing with more than two social networks, we can to incorporate this kind of constraint to further improve our method. We use $\mathcal{LT}$ to represent all such transitivity closures.

Logical transitivity feature is defined as a triangle feature:
\begin{equation}
q(y_k, y_{k'}, y_{k''}, \mathcal{Q}(y_k, y_{k'}, y_{k''})) =  \frac{1}{Z_{\eta}} e^{{\eta\textbf{q}(y_k, y_{k'}, y_{k''})}}
\end{equation}
where $y_k, y_{k'}, y_{k''}$ are nodes forming a logical transitivity closure. $\textbf{q}(y_k, y_{k'}, y_{k''})$ would be set to 0 if two of them are labeled true and the other one is labeled false since it violates the logical transitivity constraint, otherwise $\textbf{q}(y_k, y_{k'}, y_{k''})$ is set to 1. As can be seen from Figure \ref{pic:transitivity}, logical transitivity connects nodes from different network pairs, thus capturing the correlations across multiple social networks.

\subsubsection{Learning}
Model learning is to estimate the best parameter configuration \textbf{$\theta$} $= (\lambda, \beta, \gamma, \eta)$ that maximizes the log-likelihood objective function. According to Hammersley-Clifford theorem, we can obtain the following log-likelihood function.
\begin{equation}
\begin{aligned}
\mathcal{O}(\Theta) & = log P_{\Theta}(Y|G,C)\\
          &= \sum_{k=1}^{|\cup{C_{i,j}}|}[\sum_d \alpha_d \Phi(y_k,x_k)] \\
          &+ \sum_{(y_k, y_{k'}) \in \mathcal{CST}} \beta \textbf{g}(y_k, y_{k'})\\
          &+ \sum_{\mathcal{C}_l \in \mathcal{MC}} [\sum_{y_k \in \mathcal{C}_l} \gamma \textbf{h}(y_k, \mathcal{C}_l)]\\
          &+ \sum_{(y_k, y_{k'}, y_{k''}) \in \mathcal{LT}} \eta \textbf{q}(y_k, y_{k'}, y_{k''}))
           - log Z
\end{aligned}
\end{equation}
where $Z = Z_{\lambda}Z_{\beta}Z_{\gamma}Z_{\eta}$ is a normalization factor.

For representation simplicity, we concatenate all feature functions for variable $y_k$ as 
\begin{equation}
\Psi(y_k) = ({\Phi(y_k,x_k)}^T,{\textbf{g}(y_k,y_{k'})}^T,{\textbf{h}(y_k,\mathcal{C}_l)}^T,{\textbf{q}(y_k,y_{k'},y_{k''}})
\end{equation} 
We adopt gradient decent method to maximize the objective function. The gradient can be calculated as following
\begin{equation}\begin{aligned}
\frac{\partial \mathcal{O}(\theta)}{\partial \theta}  & = E{[\Psi(y_k)]}  - E_{p_{\theta}(Y|G,C)}{[\Psi(y_k)]} \label{objective}
\end{aligned}\end{equation}
The term $E{[\Psi(y_k)]}$ in Eq.\ref{objective} is easy to calculate, but the second term is intractable to directly calculate. In this work, we use Loop Belief Propagation to approximate the marginal probability $p(y_i|\theta)$. Hence the gradient is obtained by summing over all nodes.

\begin{algorithm}[t]
\label{alg:learning}
\caption{Learning algorithm.}
\small
\SetLine \KwIn{Observed social networks $G$ with candidate pairs $C$, and the learning rate $\eta$;}
\KwOut{learned parameters $\mathbf{\theta}$;}
$\theta\leftarrow\mathbf{0}$\;

\Repeat{converge}{
    \ForEach{$v_{q}\in Q$ and $q$}{
     //Initialization\;
    $L \leftarrow$ initialization list\;

Factor graph $FG \leftarrow BuildFactorGraph(C)$\;

// Learn the parameter $\mathbf{\theta}$ for factor graph model\;

Calculate $E_{p_{\theta}(Y|G,C)}{[\Psi(y_k)]}$;

\ForEach{$\theta_i\in \mathbf{\theta}$} {
    Calculate gradient $\nabla_i$ according to Eq. 8\;
    Update $\theta^{new} = \theta^{old} + \eta \cdot \nabla_i$\;
	}
}
}

\normalsize
\end{algorithm}

\subsubsection{Inference}
Based on the learned parameters $\theta$, unknown labels can be predicted by finding the optimal configuration of $Y$ than maximizes the joint probability in Equation (3), that is,
\begin{equation}
Y^* = argmax_{Y_\Theta(Y|G,C)}
\end{equation}
Utilizing marginal probabilities $p_i|Y, G$ estimated by LBP, whether a node is an overlapping user pair is labeled according to whether it has the largest marginal probability of being labeled as 1 among all the nodes constrained by subgraph constraint factor. The higher the probability is, the more confidence we have in its correctness.
For inference, we use the max-sum algorithm (the max version of Eqs. 8 and 9) to find the values of Y that maximizes the likelihood. This max-sum algorithm is similar with the sum-product algorithm, except for the message passing functions, which calculate the message according to max instead of sum.

\subsection{Constructing the Social Graph}

{\bf Adaptive Data Crawling Strategy}
Due to the highly dynamic characteristic and the huge data scale, it is almost impossible for us to obtain the entire social network. In our framework, the data are feeding in incrementally by a set of ad-hoc social network crawlers. In traditional social network crawling task, an often used technique is breath-first search which iteratively obtains data of nodes and their neighbors. However such a straight forward crawling strategy is not suitable for social network integration task since it requires overlap of different social networks. Moreover, as we discussed above, friends reveals our identities, it is much easier to determine whether to align a pair of accounts when most of it's neighbors has already been aligned. Thus, we prefer to obtain the accounts close to the aligned pairs first. 

In the framework, we store the accounts waiting to be crawled in a priority queue. After each of time of alignment phase, we iterate all the aligned pairs, for each pair $c_{i,j} = (v_i^{k_1}, v_j^{k_2})$, we give one "credit" score to each of their neighbors $NB_i(v_i^{k_1})$ and $NB_j(v_j^{k_2})$. The priority is determined by the received credit of each account. Hence, we can avoid low overlap rate between networks giving return by blind BFS crawling.

{\bf Global Social Graph Construction}
With the alignment output by the probabilistic model, we can finally construct the global social graph. Follow the definition given by Section 2 that global social graph $G_0 = (V_0, E_0)$ is a multi-graph where each node $v_0^k \in V_0$ corresponding to a set of nodes on different networks belonging to an individual natural person. We first group associated aligned pairs into a set, where relation is defined as two candidate pairs sharing a node(eg. $(v_1^1,v_2^1)$ and $(v_1^1,v_3^1)$). For each set, we create a node $v_0^k$ in $V_0$ where $v_0^k$ corresponds to all the nodes within the set. We iteratively incorporate edge for the nodes from each social network and create a new node for any node don't have correspondence in $G_0$. Base on the constructed global social graph, we developed an integrated entity search application which users can search across different social networks. We won't discuss this part of the work in this paper.

%% file: exp.tex
\section{Experimental Results}
In this section, we first introduce the data sets for experiments, and then present experimental results as well as empirical analysis.

\begin{figure*}[t]\centering
\subfigure[ArnetMiner \& Linkedin]{
\includegraphics[width = 5.5 cm]{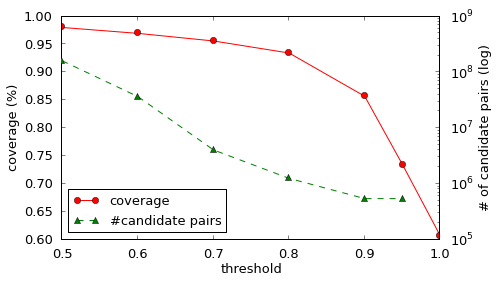}
}
\subfigure[ArnetMiner \& Videolectures]{
\includegraphics[width = 5.5 cm]{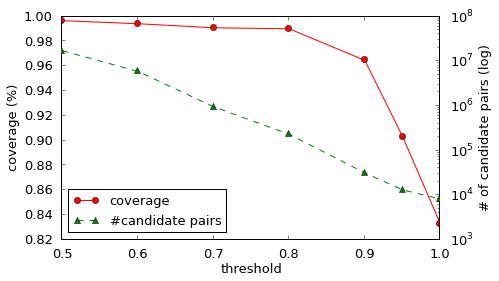}
}
\subfigure[Linkedin \& Videolectures]{
\includegraphics[width = 5.5 cm]{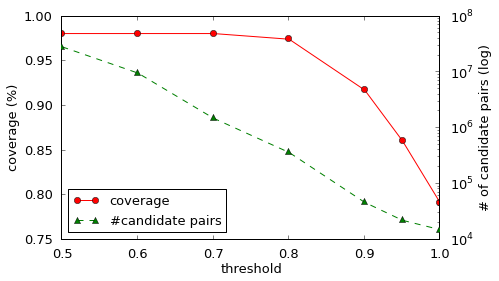}
}
\caption{Coverage rate and size of the generated candidate pair set w.r.t threshold}
\label{pic:trade-off}
\end{figure*}

\subsection{Experiment Setup}

\textbf{Candidate Pairs Generation} We perform candidate pairs generation using the method explained in Section 4. Recall in Section 4 we mentioned the preferred candidate generation method returns candidate pair set with relatively reasonable size meanwhile covers as much corresponding pairs as possible. Hence, to select an appropriate threshold, we need to make a trade off between these two factor. We evaluate the quality of candidate generation on the labeled data. Figure \label{pic:trade-off} shows the candidate coverage rate and the size of the generated candidate pair set with respected to threshold between each pair of networks. In Figure \label{pic:trade-off} we can see that the size of candidate size in growing exponentially with the decreasing of threshold. We choose a threshold 0.8 which yields high coverage rate(higher than 98\% between ArnetMiner \& Videolectures and Linkedin \& Videolectures and close to $95\%$ between ArnetMiner \& Linkedin) and reasonable candidate set size.

\textbf{Comparison Methods} 
We compare the following methods with our probabilistic model at corresponding pair alignment phase.
\begin{itemize} 
	\item Strict Name Matching (SNM). This method only uses names to estimate the correctness of a given candidate pair. For each candidate pair, it predicts positive if and only if the two associated accounts have the exact same name.
	\item Unique Name Matching (UNM). According to the discussion in Section 4.2.1, this method only returns positive for the candidate pairs with the exact same name $\mathcal{N}$ meanwhile the uniqueness of the name $\mathcal{U}(\mathcal{N}) > 20$.
	\item Logistic Regression(LR). This method incorporates all the local features discussed in section 4.2 to train a classification model and the employs the classification model to predict the correctness of   candidate pairs.
    \item Conditional Random Field (CRF). It trains a conditional random field \cite{lafferty2001conditional} with the local features associated with each candidate pair and structural correlations between candidate pairs, where correlations are indicated by the potential persistent social across different social network introduced in Section 4.2.
	\item Our Method. The proposed approach in the paper, which leverages the local features as well as the persistence of social tie, one-to-one mapping constrain and logical transitivity constrain to make the prediction.
\end{itemize}

\textbf{Evaluation Measure}
We evaluate the performance of each method in terms of precision, recall and F1-score.

\subsection{Performance Analysis}
We compare the performance of the five method for corresponding pairs alignment task. Note that our model is trained over all three social networks ArnetMiner, Linkedin and Videolectures, and making prediction jointly. But for clarity, we present the experimental result for each pair of social networks both separately and combined.

Table \ref{tb:performance} shows the empirical results for integrating three social networks: ArnetMiner, Linkedin and Videolectures. According to Table \ref{tb:performance}, our method significantly outperforms the baselines. The two rule-based baseline methods SNM and UNM only use username to make the prediction. SNM strictly aligns all the candidate pairs with the exact same username. For a commonly used name, it will predict all the candidate pairs sharing that name positive hence yields low precision. UNM considers the uniqueness of name and only considers those candidate pairs sharing a rarely used name as they belong to the same person, which gives a high precision, while recall suffers. However, as a result, the performance of the two simple rule-based method confirms that unique name is a strong indicator for corresponding pairs.

The improvement between CRF and LR verifies the effectiveness of persistent social ties for corresponding pairs alignment and further confirms the intuition that friends reveals our social identity. 

Finally, due to the extremely imbalance (\# negative pairs / \# positive pairs is approximately $10^3:1$) and lacking of training instance, all the baseline methods yield low precision for the alignment between Linkedin \& Videolectures. By encountering logical transitivity between corresponding pairs, we can leverage the information from the other two network pairs to improve the performance, thus, our method significantly outperforms the baseline methods for aligning Linkedin \& Videolectures.

\begin{table}[t]
  \centering\small\addtolength{\tabcolsep}{-1pt}
\newcommand{\minitab}[2][l]{\begin{tabular}{#1}#2\end{tabular}}
{
\renewcommand{\arraystretch}{1.2}
    \begin{tabular}{c|c|ccc}
    \hline
    \textbf{Data Set} & \textbf{Method} & \textbf{Prec.} & \textbf{Rec.} & \textbf{F1-score}\\ \hline
    \multirow{4}{*}{\minitab[c]{ArnetMiner \\ \& \\ Linkedin}} 
    & SNM & 0.5950 & 0.6994 & 0.6430\\
    & UNM & \textbf{0.9081} & 0.5141 & 0.6565\\
    & LR & 0.6765 & \textbf{0.8719} & 0.7618\\
    & CRF & 0.8484 & 0.8315 & 0.8398\\
    & Our Method & 0.8602 & \ 0.8674 & \textbf{0.8638}\\ \hline \hline

    \multirow{4}{*}{\minitab[c]{ArnetMiner \\ \& \\ Videolectures}} 
    & SNM & 0.4656 & 0.8410 & 0.5994\\    
    & UNM & \textbf{0.9192} & 0.6034 & 0.7286\\
    & LR & 0.4063 & 0.8692 & 0.5537\\
    & CRF & 0.6919 & 0.8710 & 0.7710\\
    & Our Method & 0.7076 & \textbf{0.8946} & \textbf{0.7901}\\ \hline \hline

    \multirow{4}{*}{\minitab[c]{Linkedin \\ \& \\ Videolectures}} 
    & SNM & 0.1175 & 0.8222 & 0.2055\\    
    & UNM & 0.3045 & 0.5889 & 0.4015\\
    & LR & 0.1701 & 0.7333 & 0.2762\\
    & CRF &0.3748 &0.7990 &0.5102\\
    & Our Method & \textbf{0.5012} & \textbf{0.8667} & \textbf{0.6531}\\ \hline \hline

    \multirow{4}{*}{\minitab[c]{All together}} 
    & SNM & 0.4859 & 0.7522 & 0.5904\\
    & UNM & \textbf{0.8641} & 0.5473 & 0.6702\\
    & LR & 0.5209 & 0.8673 & 0.6508\\
    & CRF & 0.6449 & 0.8041 & 0.7158\\
    & Our Method & 0.7096 & \textbf{0.8762} & \textbf{0.7841}\\ \hline \hline

  \end{tabular}
  }
  \caption{
\label{tb:performance} Performance comparison of different
methods for corresponding pair alignment task. The result are presenting combined and separately for the three network pairs .
 \normalsize
}
\end{table}

%% file: related.tex
\section{Related Works}
\label{sec:related}

\hide{
Issues with combining different data sources has existed for a long time. One popular solution is implemented based on data warehousing. The warehouse system \emph{extracts}, \emph{transforms}, and \emph{loads} data from heterogeneous sources into a single view schema so data becomes compatible with each other.
}
Due to the wide potential applications, user identification across multiple social networks has attracted more and more attention recently. Zafarani et al. \cite{zafarani:connecting} first addressed the similar problem and proposed a search based attracting. Zafarani et al. [25] takes a behavioral modeling approach models the behavioral pattern of users when selecting username, hence to further determine whether two usernames belong to the same person. Iofciu et al. \cite{iofciu:identifying} investigates the user linking problem in within a social tagging system. Liu et al. \cite{liu2013what} leverages rare username to conduct training instance for user linking task. Zhang et al. \cite{kong2013inferring} formulates the problem as a anchor link prediction task and leverages temperal and spatio information to make the prediction.

From a different angle, network integration could lead to potential leakage of privacy. A related field is called social network de-anonymization \cite{backstrom:wherefore,narayanan:de-anonymizing}. These works mainly focused on recovering users' identity from a masked social network. In [2]\cite{backstrom:wherefore}, Backstrom et al. present such process where one can identify individuals in these anonymized networks by either manipulating networks before they are nonymized or by having prior knowledge about certain anonymized nodes. Narayanan et al. \cite{narayanan:de-anonymizing} attempt to de-anonymizing two social networks based only on the network topology.

Network alignment has been studied in many works, which preforms matching between two undirected graphs. Its task is to maximize the number of "overlapped" edges given two undirected graph $G_1 = (V_1, E_1)$ and $G_2 = (V_2, E_2)$ and a set $L$ containing the feasible matching between $V_1$ and $V_2$. Due to the NP-hard nature of this definition, many different approaches are adopted to relax this constraint, such as Linear Progarmming\cite{klau2009new}, Belief Propagation\cite{bayati2009message} and so on. There also have been some successful applications under this framework, for instance, finding common pathways in biological networks \cite{singh2008global,singh2007pairwise} and ontology alignment between Citeseer papers and DBLP papers\cite{hu2008matching}.

Entity disambiguation or entity resolution is also a widely studied problem. Many existing works use Wikipedia for named entities as well as a source for entity disambiguation. Bunescu et al.\cite{bunescu2006using} exploited the Wikipedia (entity pages, redirection pages, categories, and hyperlinks) and built a context-article cosine similarity model. Cucerzan et al. \cite{cucerzan2007large} presented a large scale named entity disambiguation system that also employed a huge amount of information extracted from Wikipedia. Other works use DBpedia to construct a graph of linked data entities as a way of disambiguating named entities. Instead of considering only ontology and taxonomy\cite{gentile2010semantic,hassell2006ontology}, there are attempts to employ all the information from knowledge bases \cite{kataria2011entity, sen2012collective}. These works adapted topical models for performing entity disambiguation that considered the context of every word as well as co-occurrence patterns among entities.

Our work is also closely related to relationship mining in social network analysis. One research topic is to predict unknown link in social networks.Liben-Nowell et al. \cite{liben2007link} study the unsupervised methods for link prediction. Backstrom et al.\cite{backstrom2011supervised} proposed a supervised random walk algorithm to estimate the strength of social link. Leskovec et al. \cite{leskovec2010predicting} employed a logistic regression model to predict positive and negative links in online social networks. Wenbin et al.\cite{tang2011learning} propose a Partially-labeled Pairwise Factor Graph Model for learning to infer the type of social ties.

%% file: conclusion.tex
\section{Conclusions and future work}
\label{sec:conclusion}

In this paper, we study the problem of social network integration. We engaged an in-depth analysis across three social networks, Linkedin, Videolectures and ArnetMiner and addressed what reveals users' social identity, whether the social factors consistent across different social networks and how we can leverage this information to perform integration. Base on the observations, we proposed a unified framework for the social network integration task. It crawls data from multiple social networks and further discovers accounts corresponding to the same real person from the obtained networks. We use a probabilistic model to determine such correspondence, it incorporates features like the consistency of social status and social ties across different, as well as one-to-one mapping constraint and logical transitivity to jointly make the prediction. The empirical study shows that our framework can effectively integrate the three social networks which confirms the feasibility of the problem. 

The future work falls into two parts. 1) In this work, we are dealing with real name social network, hence we can take the advantages of using name similarity to perform candidate generation in order to shrink the search space. In anonymous social networks, such a process wouldn't be so straightforward, though we can still use username as a start point\cite{liu2013what,zafaraniconnecting}. 2) With the integrated global social graph, we want to further investigate several interesting topics such as information diffusion and user behavior pattern across different social networks that only feasible on the global social graph.